# PCPDTBT nanostructures and PCPDTBT:PC$_{71}$BM nanocomposite synthesize via modified polymer-melt technique

Muhamad Doris, Khaulah Sulaiman and Azzuliani Supangat

**Abstract**

PCPDTBT nanostructures have been synthesized via template-assisted method and polymer-melt technique. The morphological, optical and structural properties of the PCPDTBT have been investigated. Melting polymer was used as a driving force to infiltrate the Anodic Aluminum Oxide (AAO) template in which the applied temperatures are 200, 250, 300, 350, 400 and 450 Celsius respectively. The melting time duration is set constant for 30 minutes for each of those melting temperatures. Nanowires constructions have been produced at all melting temperature however based the morphological investigation, only the first two melting temperatures, 200 and 250, that can serve the structural without any broken noticed on the Raman measurement. The optical and structural properties are also confirm this fact.

# Background

# Material and methods

The primary element material in this study is a conjugated polymer Poly[2,6-(4,4-bis-(2-ethylhexyl) -4H-cyclopenta [2,1-b;3,4-b′] dithiophene)-alt-4,7(2,1,3-benzothiadiazole)] (PCPDTBT) that was purchased from Sigma-Aldrich, and its chemical structure is shown in Figure 1(a). 5 mg of PCPDTBT was dissolved in 1 ml of chloroform to produce 5 mg/ml concentration of a solution and stirred overnight without any additional purification. To construct a nanocomposite, a second material, [6,6]-phenyl $C_{71}$ butyric acid methyl ester ($PC_{71}BM$), Figure 1(b), was incorporated with PCPDTBT to produce PN-type nanocomposite.

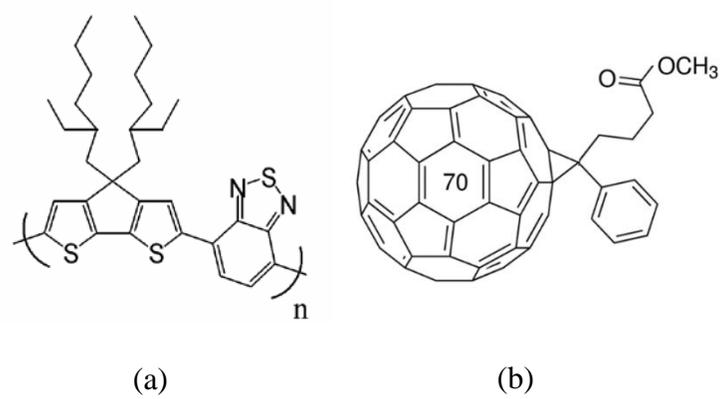

(a)                          (b)

Figure 1 Molecular structure of (a) Donor material, PCPDTBT (Cesare Soci et al., 2006), and (b) Acceptor material, PC71BM.

Inside a furnace chamber, a thick-polymer layer is placed on top of an AAO template nanopores

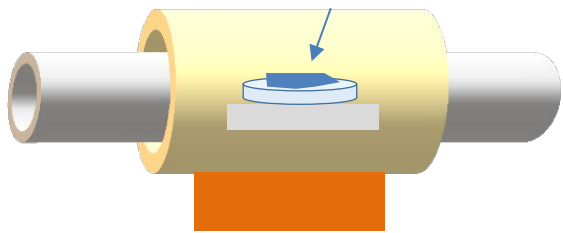

**Figure 2:** An illustration of the polymer melting process of a thick polymer layer inside a furnace chamber.

Figure 3 reveals an entire illustration of the polymer-melt method to construct the PCPDTBT nanostructures (P-type) and the PCPDTBT:PC$_{71}$BM nanocomposite (PN-type). The P-type fabrication follows step-A, while to construct the PN-type nanocomposite, the P-type needs to be combined with a second technique, spin-coating, which denoted as step-B. The P-type sample preparation was commenced by drop-casting a small amount of polymer solution onto a glass substrate and was kept for sometimes to dry it in a slow process gradually. Once the dilute polymer layer gets dry, the attached thick polymer layer was soaked into NaOH solution to peel off the polymer layer from the glass substrate.

The thick polymer layer was then put carefully on top of an empty AAO template to be melted for 30 minutes. The AAO template was then injected via melting polymer during the heating process and slowly cooled down in ambient temperature. Finally, the infiltrated AAO template was stuck up-side-down onto a copper tape and submerged within NaOH for 12 hours to dissolve the template, for an etching process, and leave behind the free-standing polymer.

For PN-type nanocomposite fabrication, the construction steps are similar as in the previous P-type production. However, in the PN-type, the second N-type material was infiltrated using a spin coating on top of the P-type. The PN-type etching process follows the same steps as been explained in the previous paragraph. Also, the etching process of an infiltrated AAO template follows the process that was conducted by Azzuliani et al. (Doris, 2017 #347).

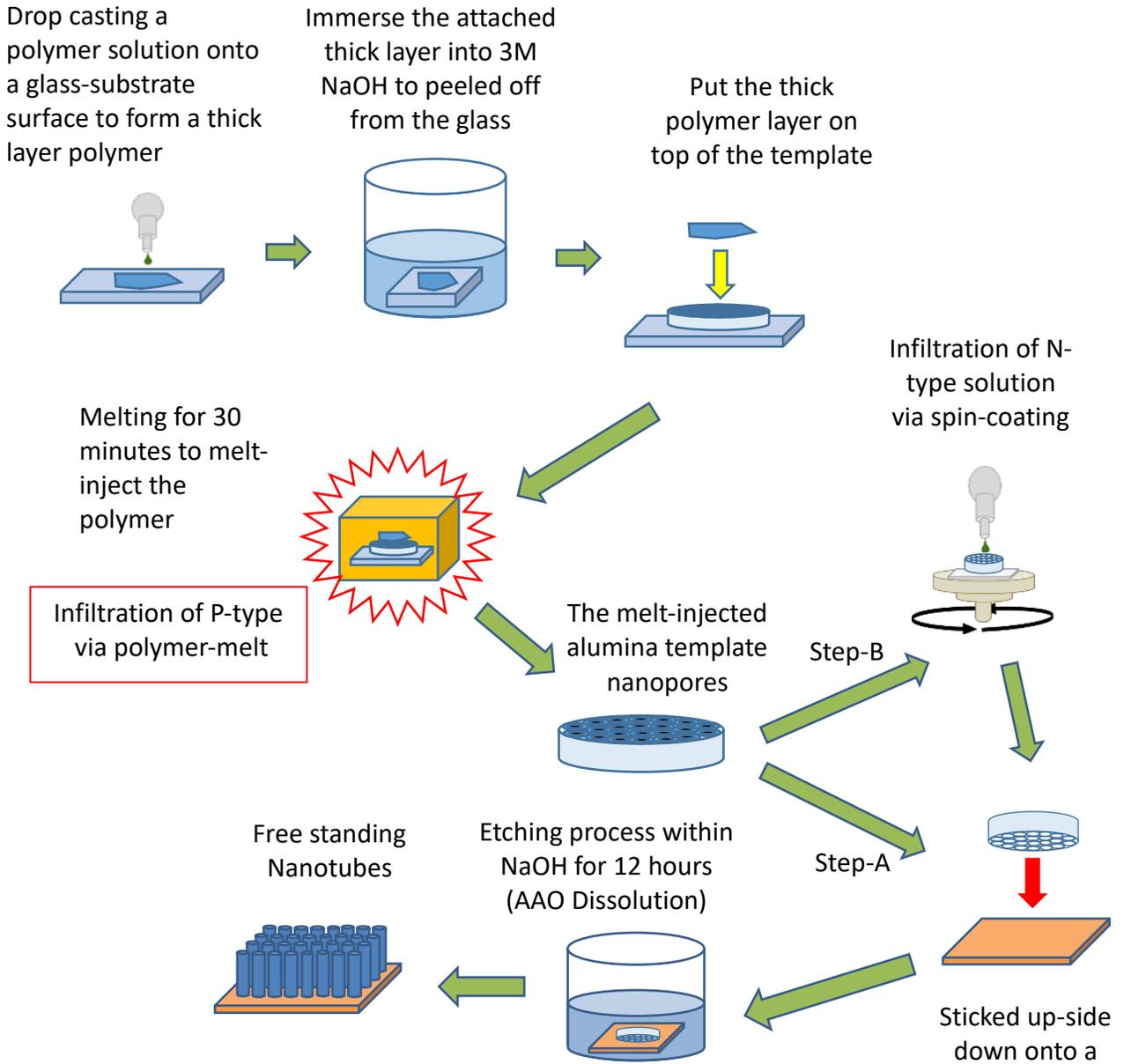

**Figure 3:** Schematic diagram of the experimental design of the polymer-melt technique to construct P-type nanostructure and PN-type nanocomposite

| Sample | Concentration (mg/ml) | Temperature (θ) | Sample label (P-type) | Sample label (PN-type) |
|---|---|---|---|---|
| 1 | 5 | 200 | PMt-200 | PNMt-200 |
| 2 | 5 | 250 | PMt-250 | PNMt-250 |
| 3 | 5 | 300 | PMt-300 | PNMt-300 |
| 4 | 5 | 350 | PMt-350 | PNMt-350 |
| 5 | 5 | 400 | PMt-400 | PNMt-400 |
| 6 | 5 | 450 | PMt-450 | PNMt-450 |

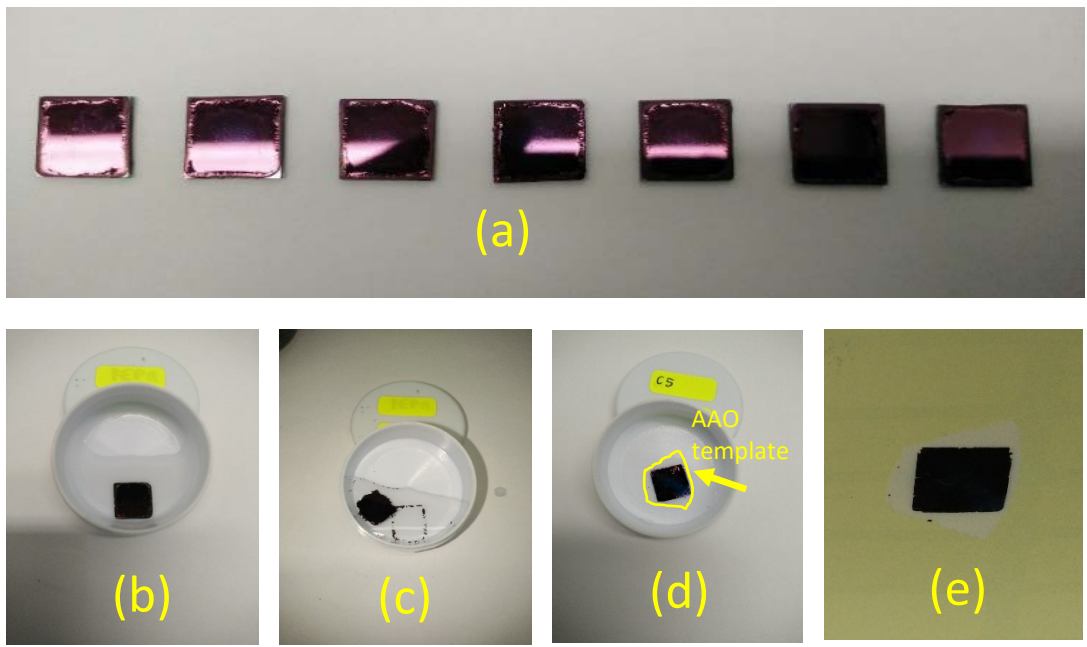

Figure 4 (a) A set of produced PCPDTBT thick layer prior immersion process into NaOH solution to be exfoliated from the glass substrate. Process of peeling off PCPDTBT thick film from a glass substrate (b); Immersion of attached polymer layer within 3 M NaOH (c); A thick polymer layer peeled off from glass surface (d); Thick polymer layer is positioned on top of AAO template after rinsed with D.I water (AAO template is shown as a yellow line), and (e) Thick polymer layer stuck on top of AAO template.

## Results and discussions

The previous study that was conducted by Azzuliani et al. has elucidated the results of nanostructures formation which produced based on the mechanical infiltration process via centrifugal force (Doris et al., 2017). In this present investigaion, thermal energy is utilized as a driving force to inject prospective material into the AAO nanopores, yet the wetting process still plays an essential role; however, without the involvement of solvent during the infiltration since the solvent is merely used to mix the intended materials to produce a thick layer polymer after the solution solidify by evaporating the solvent. Both mechanical and thermal energy altogether have ever been used to construct nanomaterials, such as nanoballs and nanowires, as what has been done by Kuo et.al in which the melting alloy at high temperature was placed in combination with a high-speed rotation of centrifugal to realize unique nanostructures whereby the AAO nanopores is still used as a primary mold (Kuo & Chao, 2005). Nevertheless, in this thesis, these two combined constructions techniques were separated into two different infiltration mechanisms instead of merging it simultaneously as one technique. Thereby, the effect of the external driving force from each source be able to differentiate and analyzed in its relationship to the properties of the revealed nanomorphology. Figure 4.12 illustrates the infiltration direction of the melt-wetting inside the nanochannel of an AAO template. Some forces that might be involved in the melt-inject process, such as cohesion, adhesion, and the gravitation. The cohesion force ($F_{coh}$) comes from interaction of molecules inside the melting polymer. The adhesion force ($F_{adh}$) is the force that exist when the polymer molecules make contact with the AAO nanochannel's wall. The third driving force that may present during the infiltration process is gravity ($F_g$). A combination of the three forces should control the nanostructures formation process inside the nanochannel other than the melting temperature itself that agitates the molecular interactions

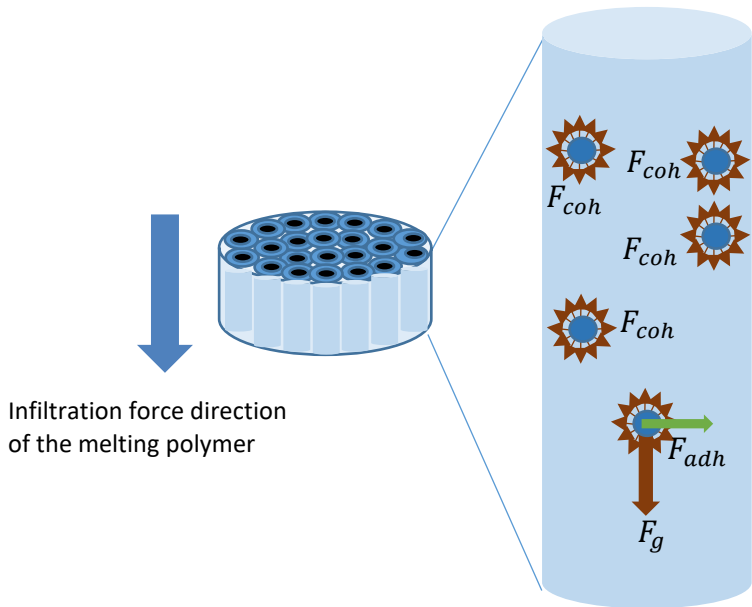

Figure 5: Illustration of polymer melting process inside the nanochannel of the AAO template.

## Morphological properties

The constructed polymer nanostructures depend on the wetting condition of the polymer. In a dilute phase, the polymer solution can yield nanotubes whereas prepare it in the melting condition lead to end-result as either nanorods or nanowires (Bordo, Schiek, & Rubahn, 2014). Figure 4.13(a)-(f) exposes PCPDTBT nanostructures construction at various melting temperatures. At $200^0$ C, the melting polymer seems to have filled in the AAO template and the revealed structures are nanowires as exhibited in Figure 4.13 (a) with groups of clusters arrangement and the wide gaps between the clusters. The length of the nanowires, ~20 µm, with an open-end tip that shown in the inset indicates a hollow tube (nanotubes). At $250^0$ C, as can be seen in Figure 4.13(b), the produced structures are quite similar to the previous one, nanowires with an open-end structure (as represented in the inset); nevertheless, its configuration is denser and longer tube, with more than ~40 µm long for every single tail.

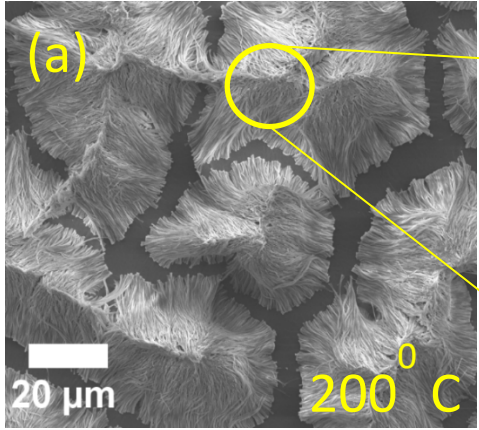
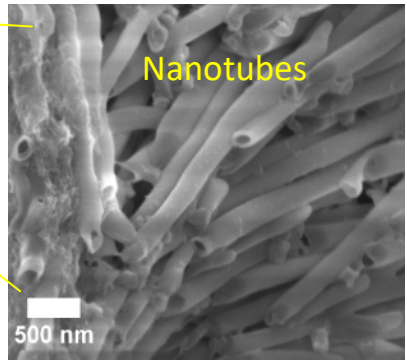

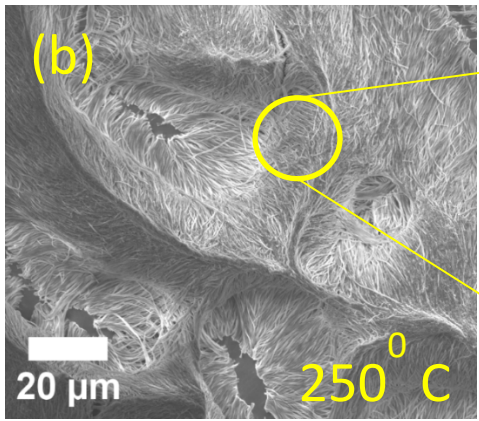
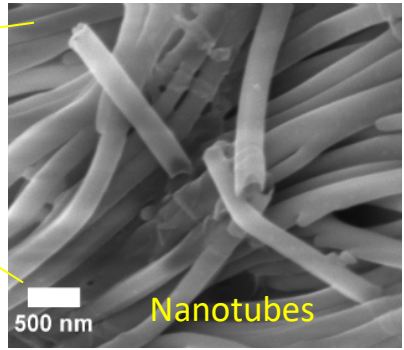

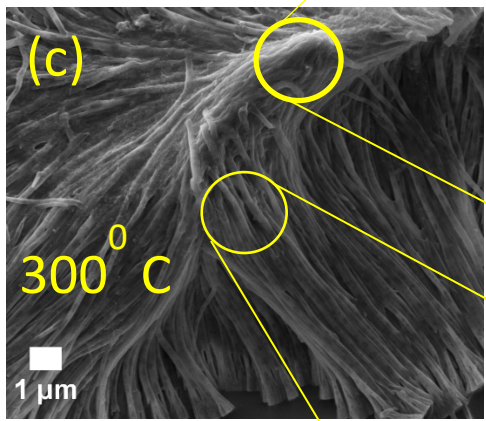
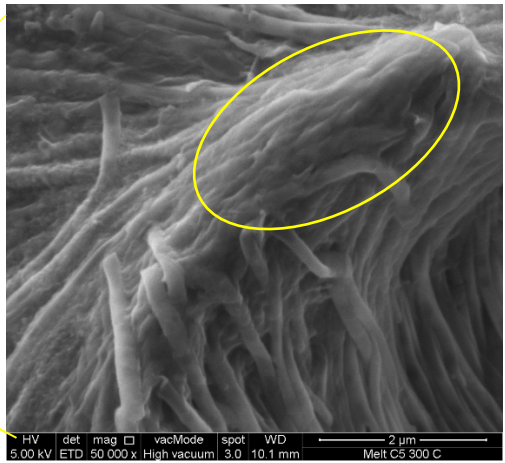
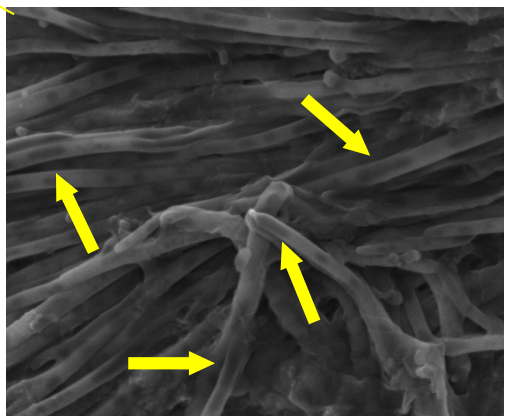

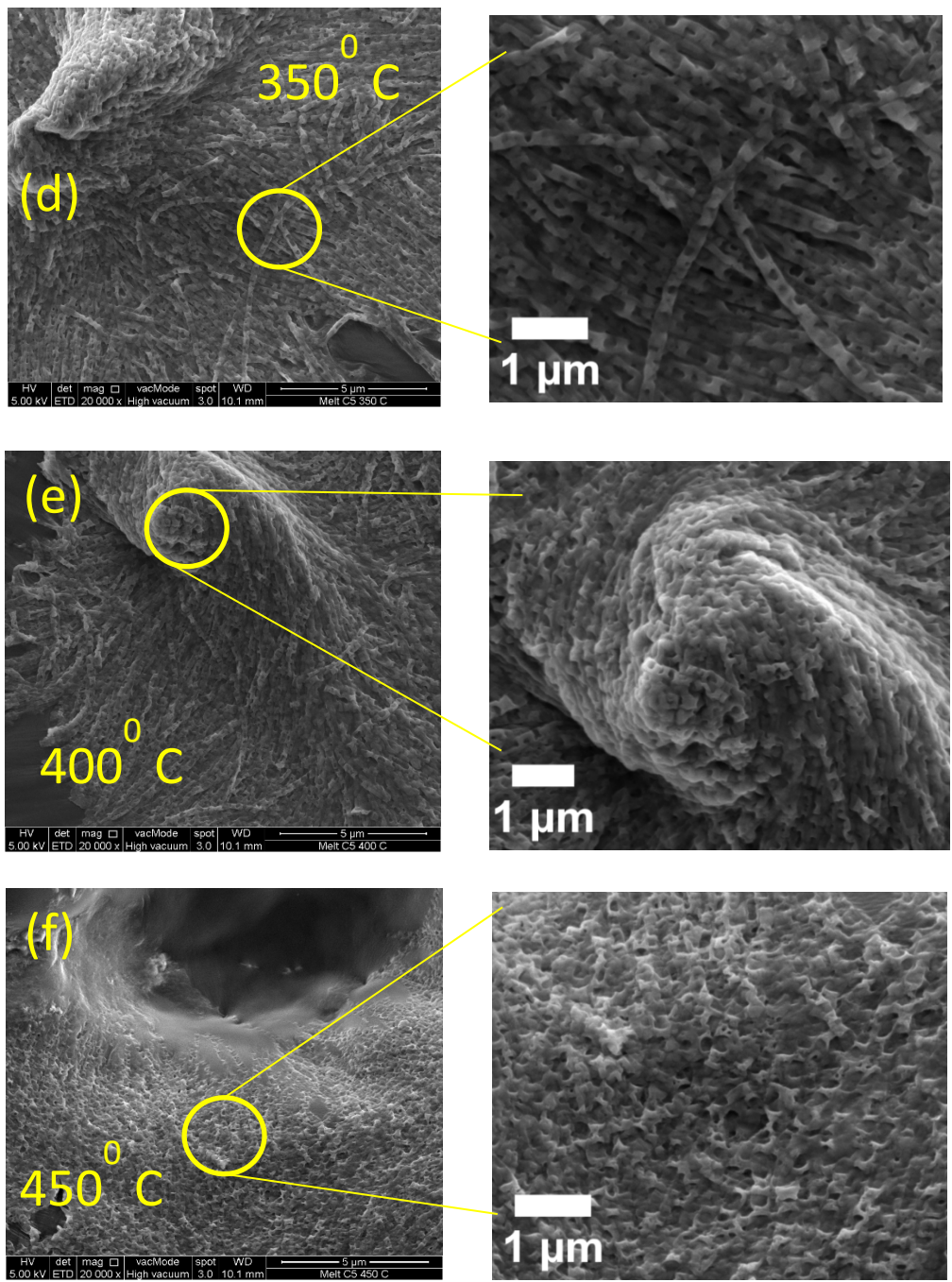

Figure 6: FESEM images of PCPDTBT nanostructures produced at various melting temperatures (200-450$^0$ C). (a) PMt-200; (b) PMt-250; (c) PMt-300; (d) PMt-350; (e) PMt-400; (f) PMt-450.

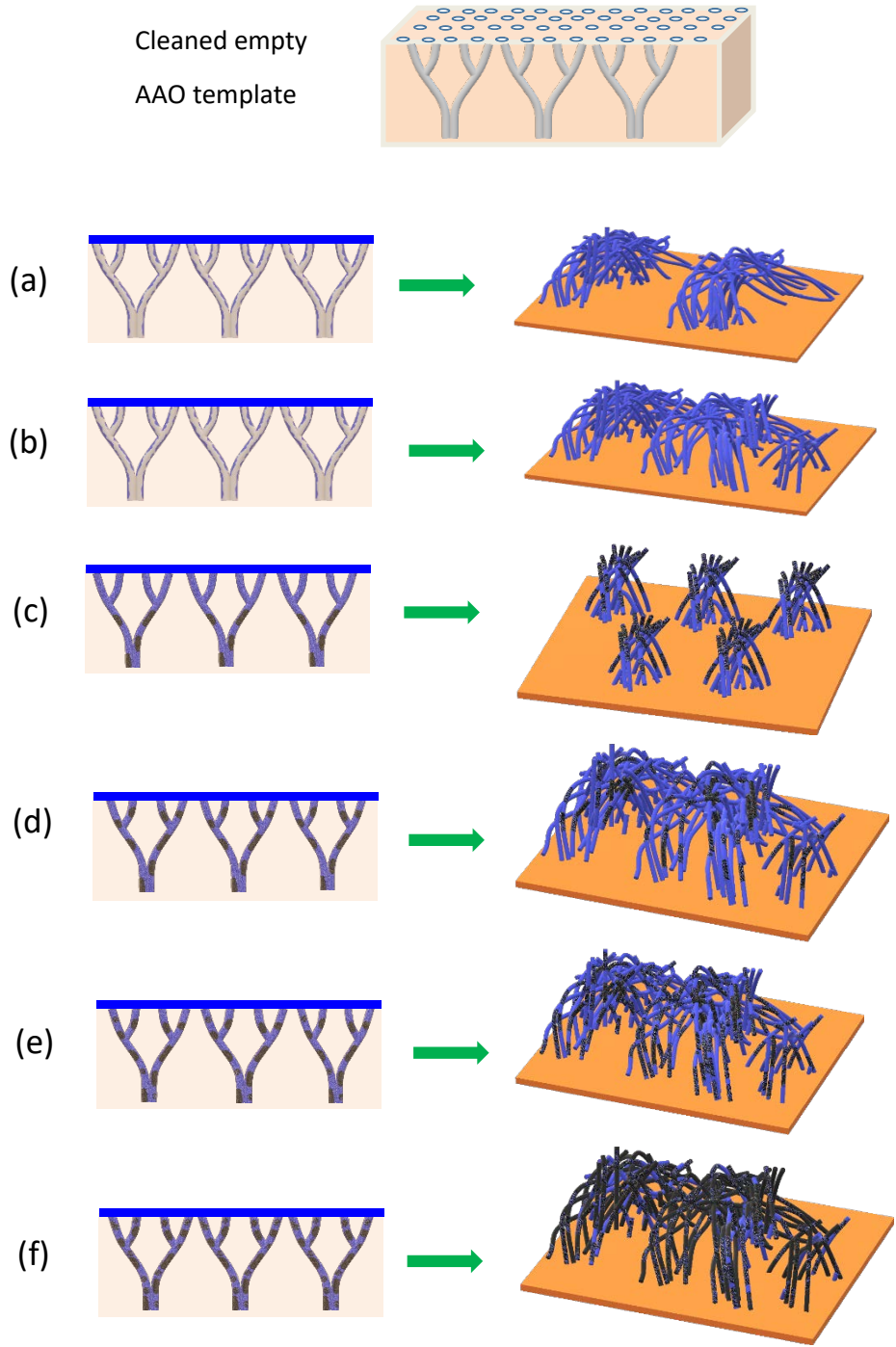

Figure 7: Schematic illustration of the proposed infiltration mechanism in the AAO nanopores via polymer-melt assisted wetting for each of melting temperature (a) PMt-200; (b) PMt-250; (c) PMt-300; (d) PMt-350; (e) PMt-400; and (f) PMt-450$^0$ C.

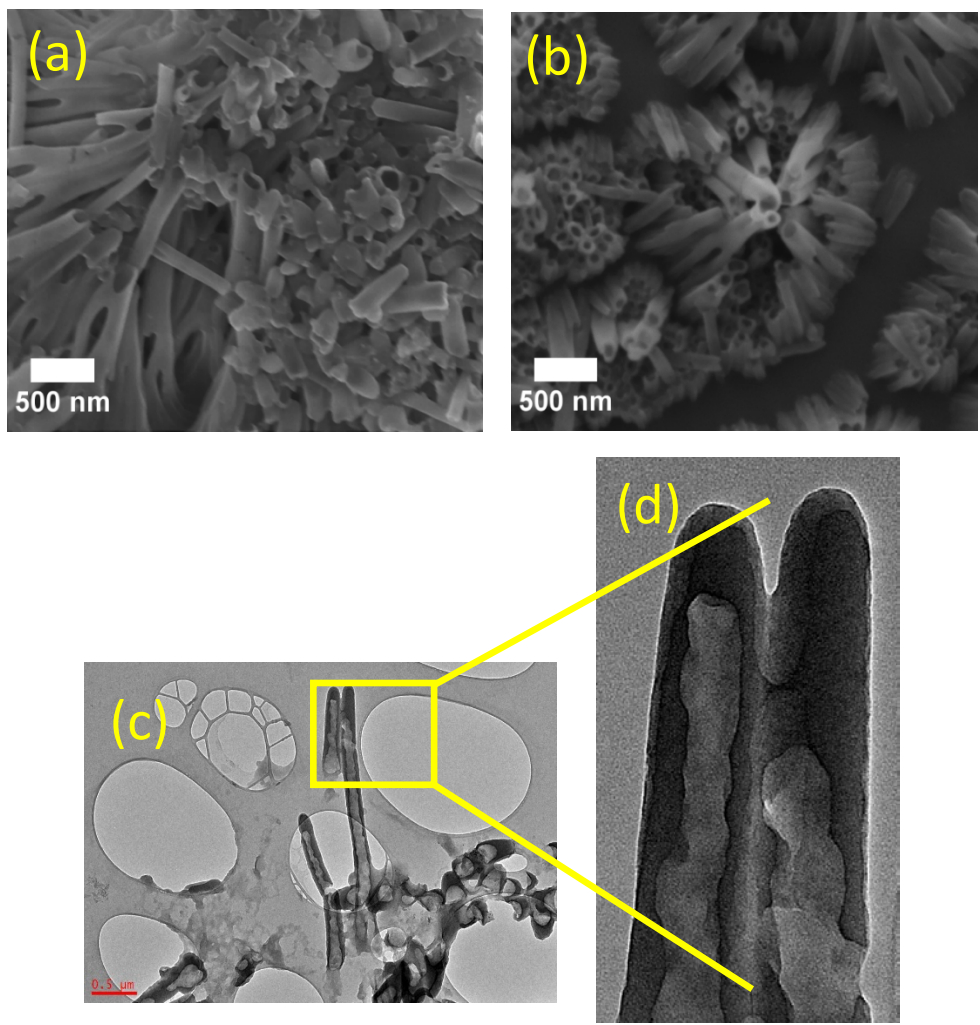

Figure 8 FESEM and TEM results of PCPDTBT:PC71BM to form PN-type nanocomposite. (a) PNMt-200 (b) PNMt-250 (c) TEM image of PNMt-250 and (d) high magnification.

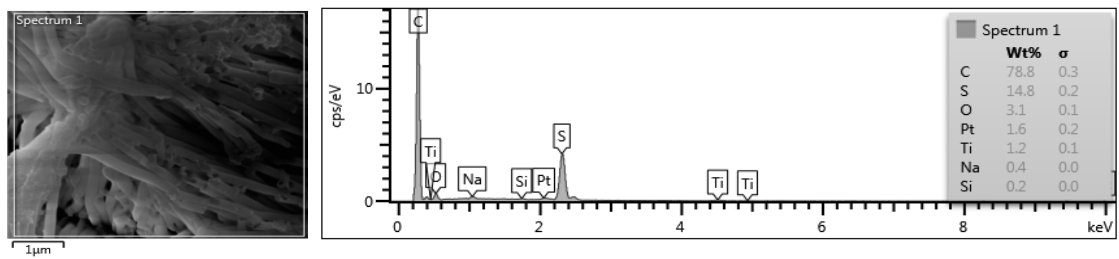

Figure 9xxxxxx

## Optical properties

Figure 10 displays the absorption and photoluminescence properties of PCPDTBT nanostructures that were produced at various melting temperatures. Generally, in a solid-state form, the absorption range characteristics of the PCPDTBT is within maxima of 350-450 nm (CPDT) and 500-850 nm (BT) (Mühlbacher et al., 2006; C. Soci et al., 2007). A higher melting temperature has made the absorption range (BT unit) of the some produced nanostructured polymer shifted to the shorter wavelength as shown in Figure 4.15(a) on samples; PMt-300, -350, -400 and -450. On the contrary, the polymer nanostructures that were fabricated at temperatures $200^0$ and $250^0$ C have a BT unit that is still in the range of 500-850 nm (PMt-200 and -250).

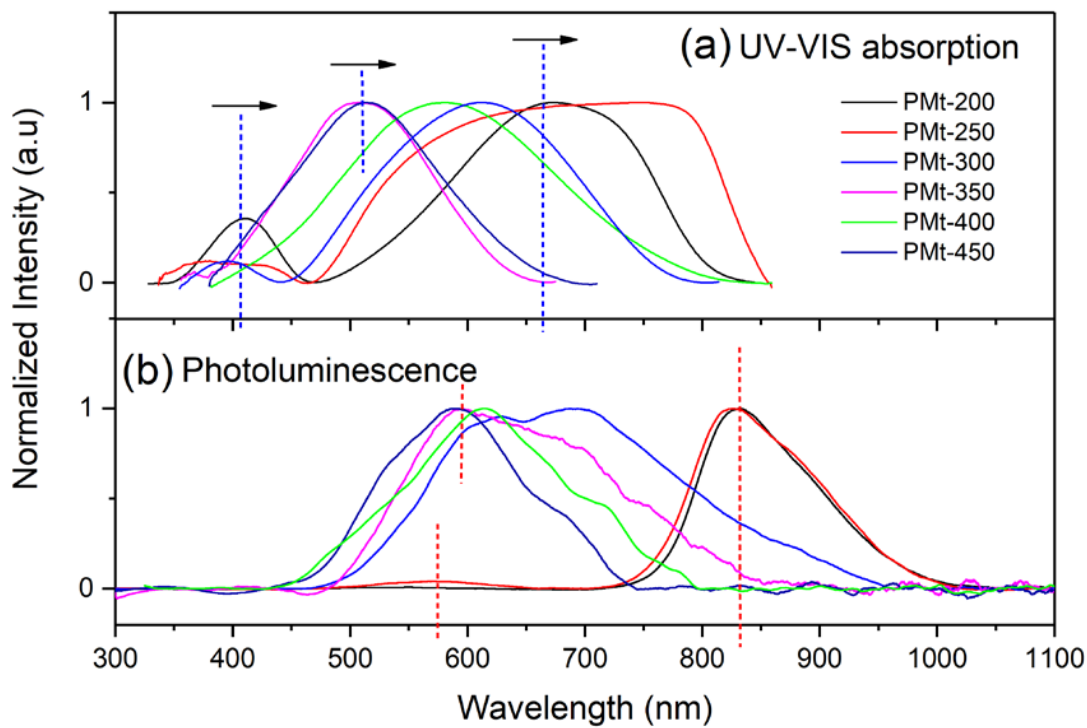

Figure 10: (a) Absorption spectra of the produced PCPDTBT nanostructures at various melting temperature and (b) the respected emission spectra.

It is observed that at temperatures $200^0$ C (PMt-200) and $250^0$ C (PMt-250) respectively are the appropriate melting temperatures for realizing PCPDTBT nanostructures. Increasing the melting temperature has shifted the absorption spectrum (BT) to the shorter wavelength which indicating the decreasing order of the polymer chain (large donor-acceptor twist angle) and conjugation length of the polymer. The disordering polymer chains have suppressed charge transfer between the CPDT and BT units (E. J. J. Martin et al., 2015).

Unarguably, it is seen the order degradation occurs on the PMt-300, PMt-350, PMt-400, and PMt-450. The shifting in the photoluminescence also follows the pattern of peaks shifting in the absorption, Stokes shift, in which the peaks are shifting to lower wavelength as the melting temperature move to a higher value, as displayed in Figure 4.15(b). Demonstrably, the degree of ordering in the polymer follows the melting temperatures that are employed. The ordering of the polymer chain is certainly also reflected in the nanomorphology as shown in the previous FESEM result in Figure 4.13(a-b) on PMt-200 and PMt-250, which observed have flawless structures compared to the other samples.

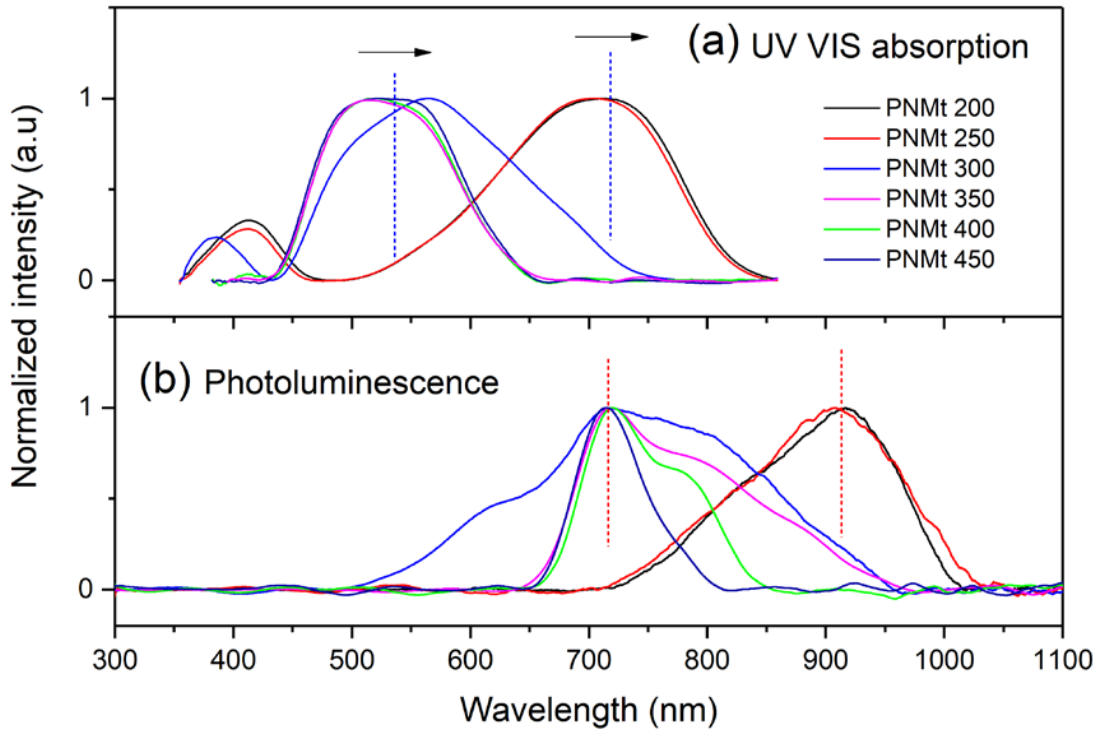

Figure 11: (a) Absorption spectra of the produced PCPDTBT:PC71BM nanocomposite and (b) the respected emission spectra.

Figure 5.10(a)&(b) diplays the UV-VIS absorption and the photoluminescence spectra of the PCPDTBT:PC$_{71}$BM respectively. It is unarguably indicating that there are only two melting temperatures, PNMt-200 and 250, provides exceptional optical response on red-shifted while the rest samples at PNMt-300, 350, 400, and 450, in contrary, are shifting to the blue shift. The revealed optical responses in the PN-type nanocomposite are similar to the previous P-type optical properties which has been investigated in Chapter 4. The BT unit absorption response lies in the range of 500-850 nm. However, the PN-type composites have redder-more shifted, maxima at 720 nm, compared to the P-type as shown in the previous chapter 4 (Figure 4.15) that stretch out in the range of 460-825 with the maxima at 660 nm. That occurs, most probably, introducing the PC$_{71}$BM into PCPDTBT has increased nanoscale phase segregation inside the PCPDTBT:PC$_{71}$BM system that might lead to a higher degree of polymer ordering (Peet et al., 2007).

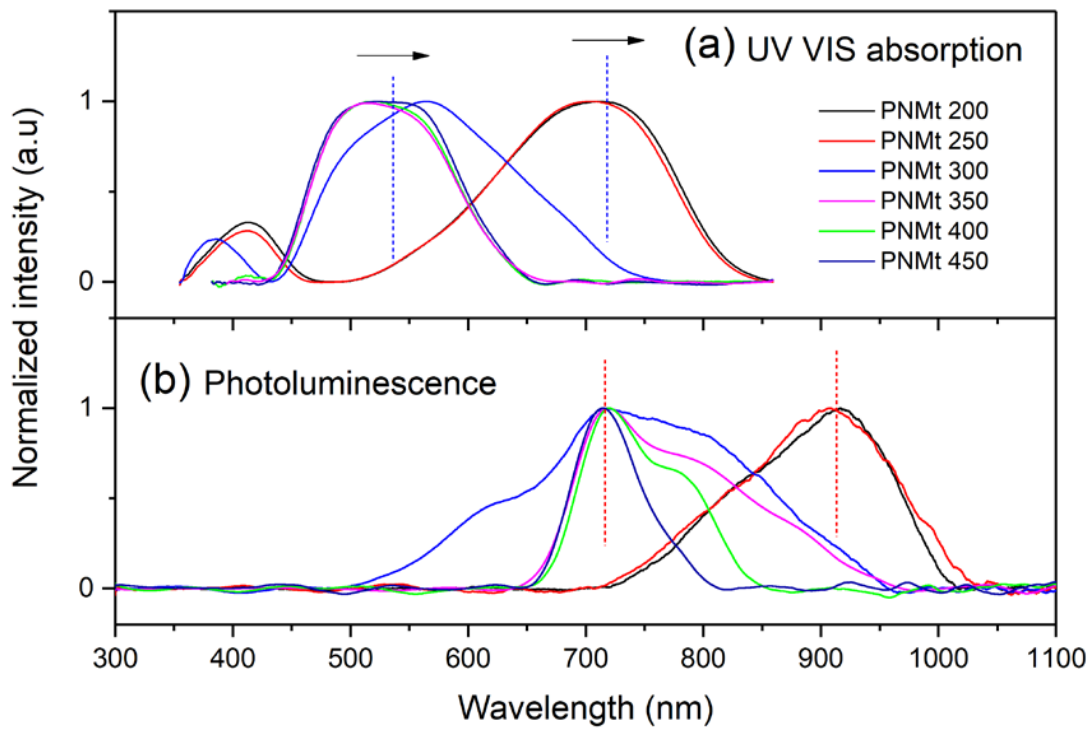

## Structural properties

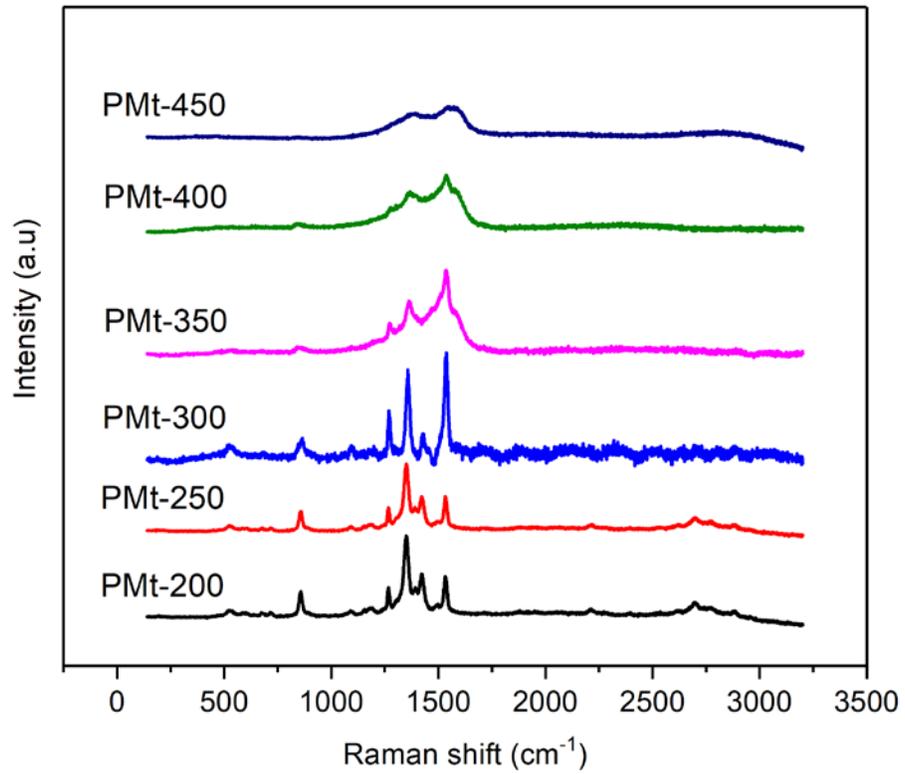

FIGURE: Raman shift measurements of the produced PCPDTBT at various melting temperatures.

TABEL: Raman mode frequencies (cm$^{-1}$) and the assignments of the P-type PCPDTBT nanostructures.

| Peak | PMt-200 | PMt-250 | PMt-300 | PMt-350 | PMt-400 | PMt-450 | Vibrational Assignments |
|---|---|---|---|---|---|---|---|
| $v_1$ | 523 | 517 | 523 | - | - | - | Thiophene: Ring deformation |
| $v_2$ | 675 | 679 | 689 | - | - | - | Out-of-plane CH deformation vibration |
| $v_3$ | 714 | 716 | 724 | - | - | - | Out-of-plane CH deformation vibration |
| $v_4$ | 858 | 857 | 865 | 841 | 840 | - | Out-of-plane CH$_2$ deformation vibration |
| $v_5$ | 1092 | 1095 | 1094 | - | - | - | CH$_3$ Rocking vibration |
| $v_6$ | 1192 | 1187 | 1198 | - | - | - | CH$_3$ Rocking vibration |
| $v_7$ | 1267 | 1268 | 1270 | 1273 | 1280 | - | BT: Symmetric In-plane (CH wag) |
| $v_8$ | 1349 | 1350 | 1357 | 1363 | 1366 | 1384 | BT: CH deformation "concert wave" (CH wag) C-H deformation (isopropyl group) CPDT: Thiophene; C=C in-plane vibration |
| $v_9$ | 1422 | 1422 | 1430 | - | - | - | CPDT: C-C stretch (Dithiophenes). CPDT: Thiophene; C=C in-plane vibration |
| $v_{10}$ | 1532 | 1532 | 1536 | 1534 | 1537 | 1542 | BT: C-C stretch (in-plane CH wag) CPDT: Thiophene; C=C in-plane vibration |
| $v_{11}$ | 2213 | 2214 | 2215 | - | - | - | C=C stretching vibration |
| $v_{12}$ | 2691 | 2697 | 2699 | - | - | - | CH stretching |

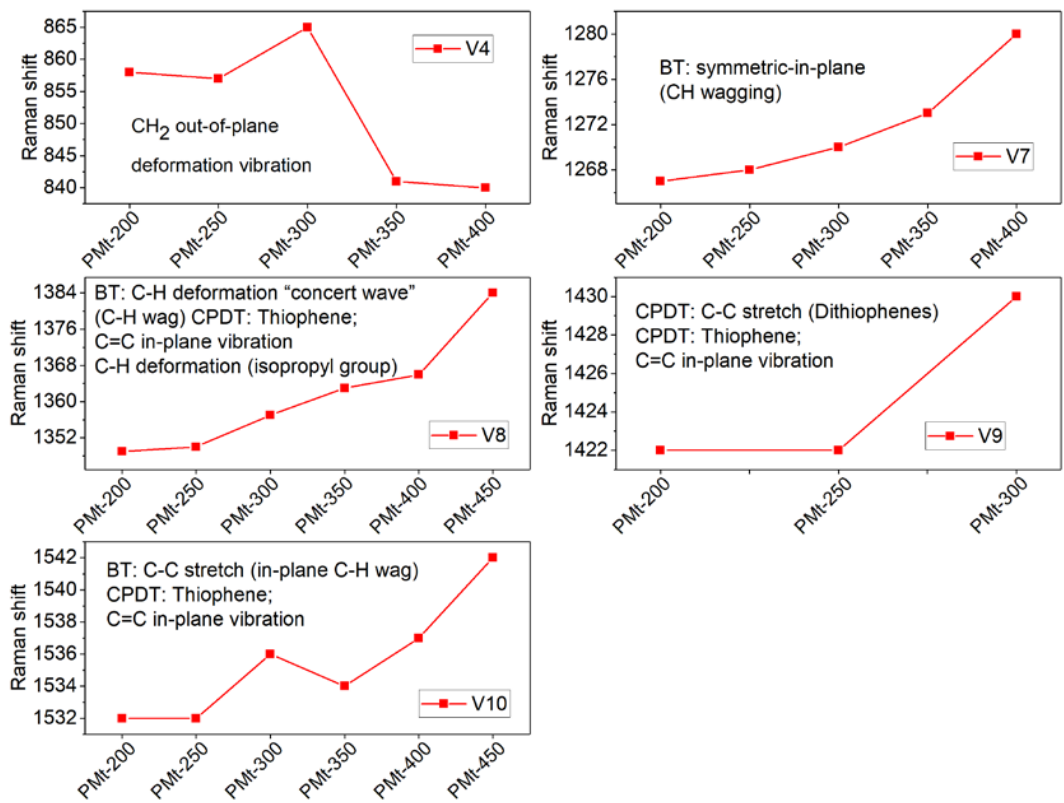

FIGURE: Summary of Raman shiftings on high intensities of the assignments derived from Table

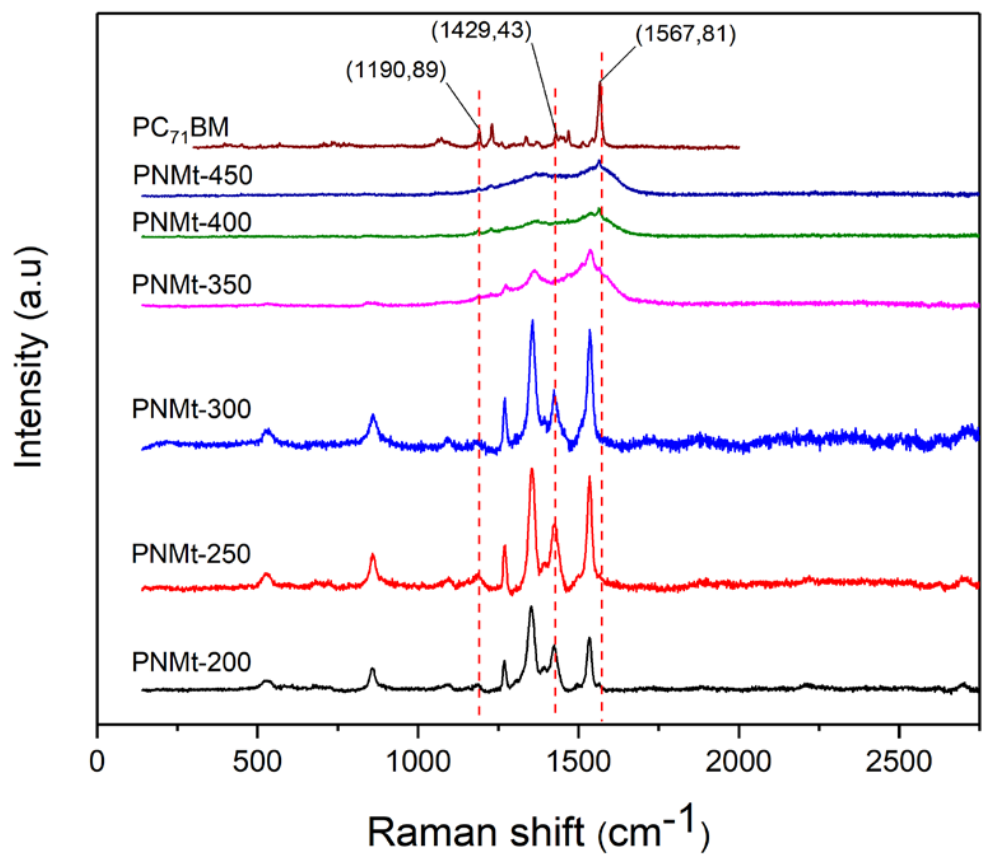

FIGURE: Raman shiftings of constructed PCPDTBT:PC$_{71}$BM

Raman mode frequencies and assignments (cm$^{-1}$) of PCPDTBT:PC$_{71}$BM nanocomposite

| Peak | PNMt-200 | PNMt-250 | PNMt-300 | PNMt-350 | PNMt-400 | PNMt-450 | Vibrational Assignments |
|---|---|---|---|---|---|---|---|
| $v_1$ | 529 | 528 | 535 | - | - | - | Thiophene: Ring deformation |
| $v_2$ | 673 | 680 | - | - | - | - | Out-of-plane CH deformation vibration |
| $v_3$ | 715 | 720 | - | - | - | - | Out-of-plane CH deformation vibration |
| $v_4$ | 857 | 858 | 859 | 845 | - | - | Out-of-plane CH$_2$ deformation vibration |
| $v_5$ | 1088 | 1095 | 1092 | 1190 | 1187 | 1189 | CH$_3$ Rocking vibration |
| $v_6$ | 1187 | 1187 | 1187 | 1227 | 1226 | 1227 | CH$_3$ Rocking vibration |
| $v_7$ | 1269 | 1271 | 1270 | 1274 | 1279 | - | BT: CH deformation (Symmetric In-plane) |
| $v_8$ | 1352 | 1355 | 1357 | 1364 | 1374 | 1372 | BT: CH deformation "concert wave" (CH wag) CPDT: Thiophene; C=C in-plane vibration & C-H deformation (isopropyl group) |
| $v_9$ | 1424 | 1423 | 1423 | - | - | - | CPDT: C-C stretch (Dithiophenes). CPDT: Thiophene; C=C in-plane vibration |
| $v_{10}$ | 1533 | 1534 | 1536 | 1536 | 1563 | 1564 | BT: C-C stretch (in-plane C-H wag) CPDT: Thiophene; C=C in-plane vibration |
| $v_{11}$ | 2213 | 2220 | 2224 | - | - | - | C≡C stretching vibration |
| $v_{12}$ | 2703 | 2694 | 2696 | - | - | - | CH stretching |

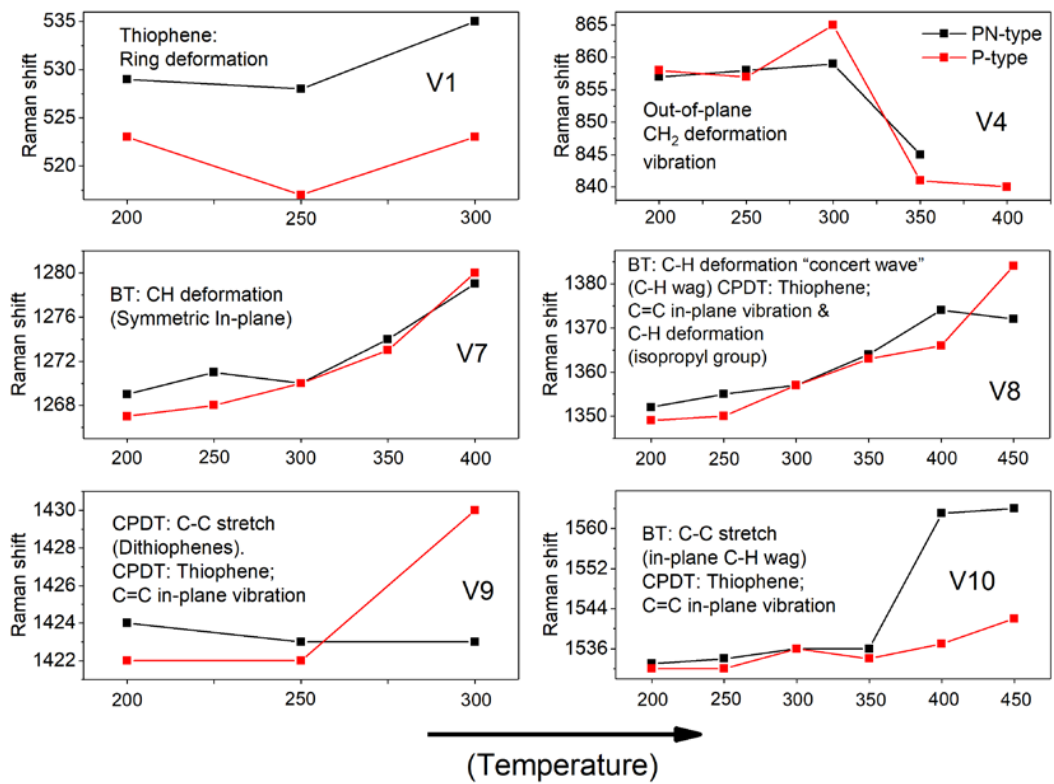

FIGURE: Summary of the Raman shifting on each of assignment based on particular high intensities derived from Table 4.2 (P-type) and Table 5.2 (PN-type).

# Conclusion

Polymer melt wetting process (Ali et al., 2015)

Wetting (Zhang et al., 2006)

There are some ways that have been applied to fabricate nanostructures by using alumina template nanopores through the wetting process either melt wetting or solution wetting. An external forces have been applied whilst a wetting process takes place. For instance, fabrication of a polymer nanostructures using melt wetting method as applying a vibration could enhance the infiltration process into the nanopores (Xie, Xu, & Yung, 2012) (Kong, Xu, Yung, Xie, & He, 2009). the surface tension and viscosity of the liquid, the size of the capillary, and the length of the channel determine the rate of liquid flow in capillary dz=dt ¼ Rg cos yc=ð4ZzÞ (Kim, Xia, & Whitesides, 1995). The viscosity of the polymer melts usually decreased with the increase of temperature. Hydrodynamics model in slit shape nanochannel (Bhadauria & Aluru, 2013)